\documentclass[twocolumn,fleqn,natbib]{svjour2}
\usepackage{amsmath}
\bibpunct{[}{]}{;}{n}{}{,} 
\smartqed  
\usepackage{graphicx}
\usepackage{mathptmx}      
\journalname{Granular Matter}
\begin{document}

\title{Close-packed granular clusters: hydrostatics and persistent Gaussian fluctuations\thanks{This research was supported by the German-Israel Foundation for
Scientific Research and Development (Grant I-795-166.10/2003) and
by MEyC and FEDER (project FIS2005-00791).
}
}

\titlerunning{Close-packed granular clusters}        

\author{Baruch Meerson \and Manuel D\'iez-Minguito \and Thomas Schwager \and Thorsten P\"{o}schel}

\authorrunning{B. Meerson et al.} 

\institute{Baruch Meerson
\at Racah Institute of Physics\\
Hebrew University of Jerusalem\\
Jerusalem 91904\\
Israel\\
\email{meerson@cc.huji.ac.il}
\and
Manuel D\'iez-Minguito \at Institute ``Carlos
I''\\ \
for Theoretical and Computational Physics\\
University of Granada\\
Spain\\
\email{mdiez@onsager.ugr.es}
\and
Thomas Schwager \at
Charit\'e\\
Augustenburger Platz 1\\
13353 Berlin\\
Germany\\
\email{thomas.schwager@charite.de}
\and
Thorsten P\"{o}schel \at
Charit\'e\\
Augustenburger Platz 1\\
13353 Berlin\\
Germany\\
\email{thorsten.poeschel@charite.de}
}

\date{Received: \today }

\maketitle

\begin{abstract}
Dense granular clusters often behave like macro-particles. We
address this interesting phenomenon in a model system of
inelastically colliding hard disks inside a circular box, driven by
a thermal wall at zero gravity. Molecular dynamics simulations show
a close-packed cluster of an almost circular shape, weakly
fluctuating in space and isolated from the driving wall by a
low-density gas. The density profile of the system agrees very well
with the azimuthally symmetric solution of granular hydrostatic
equations employing constitutive relations by Grossman \textit{et
al}, whereas the widely used Enskog-type constitutive relations show
poor accuracy. We find that fluctuations of the center of mass of
the system are Gaussian. This suggests an effective Langevin
description in terms of a macro-particle, confined by a harmonic
potential and driven by a delta-correlated noise. Surprisingly, the
fluctuations persist when increasing the number of particles in the
system.
\end{abstract}

\section{Introduction}
\label{intro}
Despite much progress in the last two decades, modeling of granular
flow remains a challenge for physicists and engineers
\cite{Campbell,Kadanoff}. The simplest case for a first-principle
modeling seems to be a granular gas, or \textit{rapid} granular
flow: a flow dominated by binary particle collisions
\cite{Brilliantov,Goldhirsch_review}. Here one can use the model of
inelastically colliding hard spheres to develop kinetic and
hydrodynamic descriptions. The simplest version of this model
involves binary collisions with a constant coefficient of normal
restitution $\varepsilon$:
\begin{equation}
  \label{eq:EMD}
  \begin{split}
    \vec{v}_i^{\,\prime} = & \vec{v}_i - \frac{1+\varepsilon}{2}\left[\left(\vec{v}_i-\vec{v}_j\right)\cdot \vec{e}_{ij}\right]\vec{e}_{ij}\,,\\
    \vec{v}_j^{\,\prime} = & \vec{v}_j + \frac{1+\varepsilon}{2}\left[\left(\vec{v}_i-\vec{v}_j\right)\cdot
    \vec{e}_{ij}\right]\vec{e}_{ij}\,,
  \end{split}
\end{equation}
where primed quantities stand for post-collisional velocities, and
$\vec{e}_{ij}\equiv
\left(\vec{r}_i-\vec{r}_j\right)/\left|\vec{r}_i-\vec{r}_j\right|$.

Assuming \textit{molecular chaos} allows to use the Boltzmann or
Enskog kinetic equation, properly generalized to account for the
inelasticity of the particle collisions. Systematic derivations of
the hydrodynamic equations from the Boltzmann or Enskog equation
\cite{Brey1,Lutsko,Sela} use an expansion in the Knudsen number
and, in some versions of  theory, in inelasticity \cite{Sela}. A
finite inelasticity immediately brings complications
\cite{Goldhirsch_review}. Correlations between particles,
developing already at a moderate inelasticity, may invalidate the
molecular chaos hypothesis \cite{correlations1,correlations3,correlations2}. The normal stress
difference and deviations of the particle velocity distribution
from the Maxwell distribution also become important for inelastic
collisions. As a result, the Navier-Stokes granular hydrodynamics
should not be expected to be quantitatively accurate beyond the
limit of small inelasticity.

A further set of difficulties for hydrodynamics arise at large densities. In
mono-disperse systems strong correlations appear, already for \textit{elastic}
hard spheres, at the disorder-order transition.  As multiple thermodynamic
phases may coexist there \cite{Chaikin}, a general continuum theory of hard
sphere fluids, which has yet to be developed, must include, in addition to the
hydrodynamic fields, an order-parameter field. Furthermore, even on a specified
branch of the thermodynamic phase diagram, we do not have first-principle
constitutive relations (CRs): the equation of state and transport coefficients.
Last but not least, kinetic theory of a finite-density gas of elastic hard
spheres \textit{in two dimensions} has notorious difficulties related to the
long-time tail in the velocity pair autocorrelation function \cite{divergence}.

It has been argued recently that discrete particle noise may play a dramatic
role in rapid granular flow \cite{Barrat,Swinney,MPSS}. A new challenge for
theory is a quantitative account of this noise. A promising approach at small or
moderate densities is ``Fluctuating Granular Hydrodynamics'': a Langevin-type
theory that takes into account the discrete particle noise by adding
delta-correlated noise terms in the momentum and energy equations, in the spirit
of the Fluctuating Hydrodynamics by Landau and Lifshitz \cite{LL}. There is a
recent progress in this area in the case of small densities \cite{Brey2}. For
high densities such a theory is presently beyond reach.

This work addresses granular hydrodynamics and fluctuations in a simple
two-dimensional granular system under conditions when existing hydrodynamic
descriptions \cite{Brey1,Lutsko,Sela} break down because of large density,
\textit{not} large inelasticity. In view of the difficulties mentioned above,
attempts of a first-principle description of hydrodynamics and fluctuations
should give way here to more practical, empiric or semi-empiric, approaches. One
such approach to a hydrodynamic (or, rather, hydrostatic, as no mean flow is
present) description was suggested in 1997 by Grossman \textit{et al.}
\cite{Grossman}. The present work puts it into a test in an extreme case when
macro-particles (granular clusters with the maximum density close to the
hexagonal close packing) form. The model system we are dealing with was first
introduced by Esipov and P\"{o}schel \cite{Poeschel}. It is an assembly of $N
\gg 1$ identical disks of mass $m$, diameter $d$ and coefficient of normal
restitution $\varepsilon$, placed inside a circular box of radius $R$ at zero
gravity. The circular wall of the box is kept at constant temperature $T_0$. We
measure, using molecular dynamics (MD) simulations, the radial density profiles
of the system, including the close-packed part. Furthermore, we solve
numerically a set of granular hydrostatic equations which employ the CRs by
Grossman \textit{et al.} and show that, in a wide range of parameters, there is
good agreement between the two. We also show that, for the same setting, the
Enskog-type CRs \cite{JR} perform poorly. Finally, we investigate in some detail
the fluctuations of the macro-particle's position, by measuring the radial
probability distribution function (PDF) of the center of mass of the system.
These fluctuations turn out to be Gaussian, which suggests an \textit{effective}
Langevin description of the system in terms of a macro-particle, confined by a
harmonic potential and driven by a white noise. Surprisingly, the fluctuations
persist as the number of particles in the system is increased.

\section{MD simulations}

\begin{figure}
\centerline{\includegraphics[width=8.5cm,clip=]{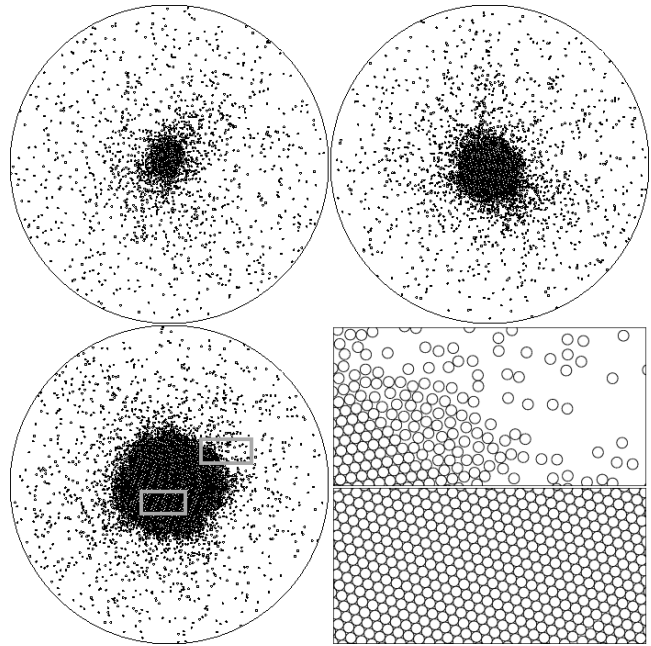}}
\caption{Snapshots of the system for $N=1716$ (top left), $2761$
(top right) and $N=5055$ (bottom left). Also shown are
magnifications of the indicated areas.} \label{fig1}
\end{figure}

We employed a standard event-driven algorithm and thermal wall
implementation \cite{Algo}. We put $m=d=T_0=1$ and fixed $R=100$ and
$\varepsilon=0.888$, while the total number of particles $N$ served
as the control parameter in the simulations. We were mostly
interested in a hydrodynamic (low Knudsen number) regime, when the
mean free path is small compared to the system size. This requires
$N \gg R/d =100$. For $N$ in the range of a few hundred, we observed
a dilute granular gas with an increased density in the center of the
box. The clustering in the center becomes more pronounced as $N$
grows. The clustering can be easily explained in the hydrodynamics
language: because of the inelastic collisions the granular
temperature goes down as one moves away from the wall toward the
center of the box. Combined with the constancy of the pressure
throughout the system, this causes an increased particle density at
the center. As $N$ increases, the particle density in the center
approaches the hexagonal close packing value $n_c = 2/(\sqrt{3}
d^2)$. Figure \ref{fig1} shows snapshots of the system for three
different, but sufficiently large, values of $N$. A perfect
hexagonal packing is apparent. Movies of these simulations
show that the cluster position fluctuates around the
center of the box, while the cluster shape fluctuates around a
circular shape.

Our diagnostics focused on two radial distributions: the number density of the
particles $n(r)$ and the PDF of a radial position of the center of mass of the
system $P(r_m)$, see below. As we are interested in steady-state distributions,
we disregarded initial transients.
%

%

\section{Hydrostatic theory}

As the fluctuations are relatively weak, it is natural to start with a purely
hydrodynamic description. In the absence of time-dependence and for a zero mean
flow this is essentially a \textit{hydrostatic} theory. It operates only with
(time-independent) granular density $n(\mathbf{r})$, temperature
$T(\mathbf{r})$ and pressure $p(\mathbf{r})$. The energy input at the thermal
wall is balanced by dissipation due to inter-particle collisions, and one can
employ the momentum and energy balance equations:
\begin{equation}
p={\rm const}\,,\,\,\, \nabla \cdot (\kappa \, \nabla T) = I \,.
\label{energy1}
\end{equation}
Here $\kappa$ is the thermal conductivity and $I$ is the rate of
energy loss by collisions. Note that the heat flux, entering the
thermal balance in Eq. \eqref{energy1}, does not include an
inelastic term, proportional to the \textit{density} gradient
\cite{Brey1,Lutsko,Sela}. In the nearly elastic limit
$1-\varepsilon \ll 1$, that we are interested in, this term can be
neglected. The boundary condition at the thermal wall is
\begin{equation}
T(r=R,\phi)=T_0\,,
\end{equation}
 where $r$ and $\phi$ are polar coordinates with
the origin at the center of the box.

To proceed from here, we need  CRs: an equation of state $p=p\,(n,T)$ and
relations for $\kappa$ and $I$ in terms of $n$ and $T$. As we attempt to
describe close-packed clusters, the standard techniques, based on the Boltzmann
or Enskog equations, are inapplicable. Grossman \textit{et al.} \cite{Grossman}
suggested a set of semi-empiric CRs in two dimensions, that are valid for all
densities, all the way to hexagonal close packing. Their approach ignores
possible phase coexistence beyond the dis\-order-order transition and assumes
that the whole system is on the thermodynamic branch extending to the hexagonal
close packing. Grossman \textit{et al.} employed free volume arguments in the
vicinity of the close packing, and suggested an interpolation between the
hexagonal-packing limit and the well-known low-density relations. The resulting
CRs \cite{Grossman} read
\begin{equation}
p= n\,T\, \frac{n_c+n}{n_c-n}\,,\,\,\,\,\,\,\kappa=\frac{\mu\,
n\,(\alpha l + d)^2 T^{1/2}}{l} \,, \label{state}
\end{equation}
and
\begin{equation}
I= \frac{\mu}{\gamma l}\, (1-\varepsilon^2) \,n \, T^{3/2}\,.
\label{sink}
\end{equation}
Here $l$
is the mean free path, which is given by an interpolation formula
\cite{Grossman}:
\begin{equation}
l=\frac{1}{\sqrt{8}n d}\, \frac{n_c-n}{n_c-an}\,,
\label{mfp}
\end{equation}
where $a= 1-(3/8)^{1/2}$. The CRs by Grossman \textit{et al.} include three
adjustable parameters of order unity: $\alpha$, $\gamma$ and $\mu$, where the
latter drops out from the steady-state problem. Grossman \textit{et al.}
determined the optimum values $\alpha=1.15$ and $\gamma=2.26$, from a comparison
between MD simulations of a system of inelastic hard disks in a rectangular box,
driven by a thermal wall, and numerical solutions of the hydrostatic equations
\eqref{energy1} in rectangular geometry. We adopted the same values of $\alpha$
and $\gamma$ for the circular geometry.

Employing Eqs.~\eqref{state} and \eqref{sink} we can reduce Eqs.~\eqref{energy1}
to a single equation for the rescaled inverse density $z(r,\phi)\equiv
n_c/n(r,\phi)$. In the rescaled coordinates $\mathbf{r}/R \to \mathbf{r}$ the
circle's radius is $1$, and the governing equation becomes
\begin{equation}
\frac{1}{r} \frac{\partial}{\partial r} \left[ r F(z)\frac{\partial z}{\partial
r}\right] +\frac{1}{r^2}\frac{\partial}{\partial \phi}\left[F(z) \frac{\partial
z}{\partial \phi}\right]=\Lambda Q(z)\,, \label{govern0}
\end{equation}
where
\begin{equation}
  \begin{split}
F(z)=&\frac{(z^2+2z-1)\left[\alpha z(z-1)+\sqrt{32/3}(z-a)\right]^2}
{(z-a)(z-1)^{1/2}z^{3/2}(z+1)^{5/2}}\,,\\
Q(z)=&\frac{(z-a) (z-1)^{1/2}}{(z+1)^{3/2} z^{1/2}}\,,
  \end{split}
\label{Q}
\end{equation}
and
\begin{equation}
\Lambda = (32/3\gamma)\,(R/d)^2\,(1-\varepsilon^2)
\end{equation}
is the hydrodynamic inelasticity parameter introduced in Ref. \cite{LMS}. As the
total number of particles $N$ is fixed, $z^{-1}(r,\phi)$ satisfies a
normalization condition:
\begin{equation}
  \int_0^{2 \pi} \,d\phi \int_0^{1} \,dr\,r\, z^{-1}(r,\phi) = \pi f\,,
  \label{normfull}
\end{equation}
where
\begin{equation}
f=\frac{\sqrt{3}}{2 \pi}N\left(\frac{d}{R}\right)^2
\end{equation}
is the average area fraction of the particles.

Azymuthally-symmetric solutions of Eq.~\eqref{govern0}, $z(r,\phi)=Z(r)$, are
described by the following ordinary differential equation:
\begin{equation}
\frac{1}{r}\frac{d}{dr}\left[r F(Z)\,\frac{dZ}{dr}\right]=\Lambda\, Q(Z)\,,
\label{ode}
\end{equation}
while the normalization condition \eqref{normfull} becomes
\begin{equation}
  \int_0^1 \, Z^{-1}(r)\,r \,dr = f/2\,.
  \label{normalization}
\end{equation}
Equations \eqref{ode} and \eqref{normalization}, together with the obvious
boundary condition
\begin{equation}
\left.\frac{dZ}{dr}\right|_{r=0} = 0\,,
\end{equation}
form a complete set. The hydrostatic density profile is completely determined by
two scaled parameters $\Lambda$ and $f$, and can be found numerically as we will
show shortly.

\begin{figure}
\centerline{\includegraphics[width=8.0cm,clip=]{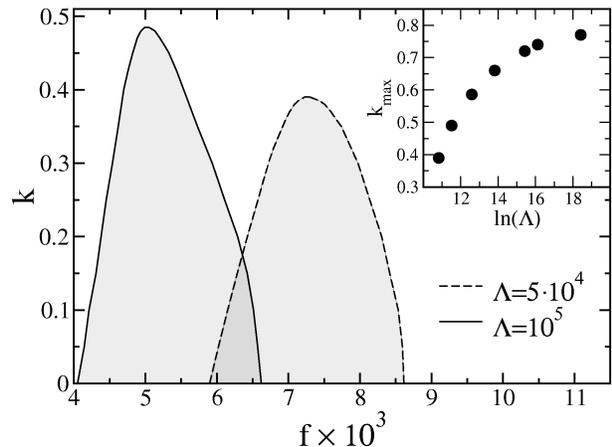}} 
\caption{The
marginal stability curves $k=k(f)$ for $\Lambda=5\times 10^{4}$ and
$\Lambda=10^{5}$, as indicated. The shaded area denotes in both cases the
(unphysical) instability region that is obtained if one ignores the quantization
of $k$: $k=1,2, \dots$. The inset shows $k_{max}$ as a function of $\ln
(\Lambda)$.} \label{fig2b}
\end{figure}

Are the azimuthally symmetric states stable with respect to small perturbations?
We performed marginal stability analysis to find out whether there are
steady-state solutions with broken azimuthal symmetry, $z(r,\phi)$ that
bifurcate continuously from an azimuthally symmetric solution $Z(r)$. Under
additional assumption that the possible \textit{instability} of the azimuthally
symmetric state is purely growing (that is, \textit{not} oscillatory), the
marginal stability analysis yields the instability borders. The marginal
stability analysis goes along the same lines as that developed for the
rectangular geometry \cite{KM,KMS,LMS,LMS2,MPSS}. Let us search a steady-state
close to an azymuthally-symmetric state:
\begin{equation}
z(r,\phi)=Z(r)+\varepsilon \; \Phi_{k}(r) \, \sin(k\, \phi)
\;,\;\;\;\;\;\varepsilon \ll 1\,.\label{eq:linear}
\end{equation}
As $z(r,\phi+2\pi) = z(r,\phi)$, $k$ must be an integer which can be chosen to
be non-negative. Substituting Eq.~(\ref{eq:linear}) into Eq.~\eqref{govern0} and
linearizing around the azimuthally symmetric state $Z(r)$, we obtain a linear
eigenvalue problem, where $k=k(f,\Lambda)$ plays the role of the eigenvalue:
\begin{equation}
\zeta_{k}^{\prime\prime}+\frac{1}{r}\zeta_{k}^{\prime}- \left[
\frac{k^2}{r^2}+\frac{\Lambda Q^{\prime}(Z)}{F(Z)}\right] \zeta_{k}=0\;,
\label{eq:eigen}
\end{equation}
where $\zeta_{k}(r)=\Phi_{k}(r)\, F(Z)$. Equation~\eqref{eq:eigen} is
complemented by the boundary conditions
\begin{equation}
\zeta_{k}(0)=0\qquad\mbox{and} \qquad\zeta_{k}(1)=0 \label{eq:eigen_bound}
\end{equation}
and can be solved numerically. Let us ignore for a moment the quantization of
the eigenvalue $k$ and, while looking for $k$, assume that it is a (positive)
\textit{real} number. In that case, a numerical solution yields, for a fixed
$\Lambda$, a curve $k=k(f)$, see Fig.~\ref{fig2b} for two examples. At a fixed
$k$, the azimuthally symmetric state is unstable within an interval of area
fractions. The foots of one such curve corresponds to the (hypothetical) case
when $k$ tends to zero (that is, the azimuthal wavelength tends to infinity).
The instability interval becomes narrower when $k$ is increased, and it shrinks
to a point at a maximum $k_{max}$, signalling that a density modulation with a
sufficiently short azimuthal wavelength should be stable for all $f$. For
example, when $\Lambda=5\times 10^4$, the marginal stability curve $k=k(f)$ has
its maximum at $k_{max}\approx 0.38$ (see Fig.~\ref{fig2b}) that is less than unity. Going back to the
physical case, where $k$ is a positive integer, we see that all the values for
$k$, determined from the eigenvalue problem, are unphysical. That is, there are
no solutions to the eigenvalue problem that would satisfy the boundary
conditions and the quantization condition $k=1,2,3 \dots$. We observed a similar
behavior for values of $\Lambda$ up to $10^8$. Though $k_{max}$ increases with
$\Lambda$, the increase is extremely slow: slower than logarithmical (see the
inset of Fig.~\ref{fig2b}).  These numerical results strongly indicate that the
azimuthally symmetric states are stable with respect to small perturbations.
This is in marked contrast with the presence of bifurcating states with broken
symmetry in similar settings of a granular gas driven by a thermal wall, but in
rectangular \cite{KM,KMS,LMS,LMS2,MPSS} and annular \cite{DM} geometries. The
absence of the bifurcating states with broken symmetry in the circular geometry
gives a natural explanation to the persistence of circle-shaped cluster shapes
as observed in our MD simulations.

\begin{figure}
\centerline{\includegraphics[width=8.5cm,clip=]{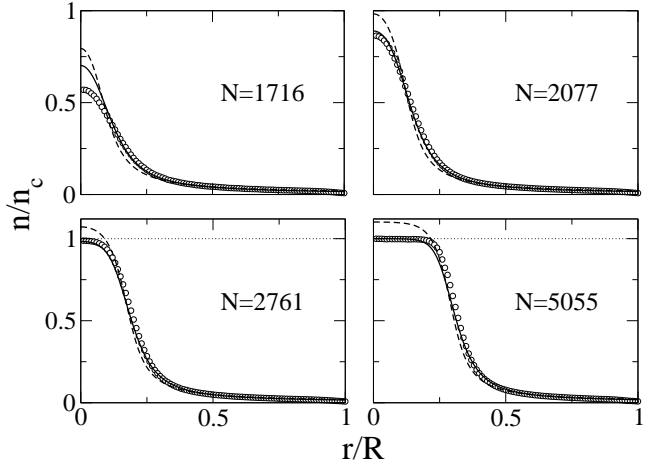}}
\caption{Scaled density $n/n_c$ versus scaled radius $r/R$ as
observed in MD (circles) and predicted by the hydrostatic theory
with the CRs by Grossman \textit{et al.} \cite{Grossman} (solid
lines) and with the Enskog-type CRs \cite{JR} (dashed lines) for
four different values of the total number of particles $N$. For the
rest of parameters please see the text. The dotted lines indicate
$n/n_c=1$.} \label{fig2}
\end{figure}

\begin{figure}
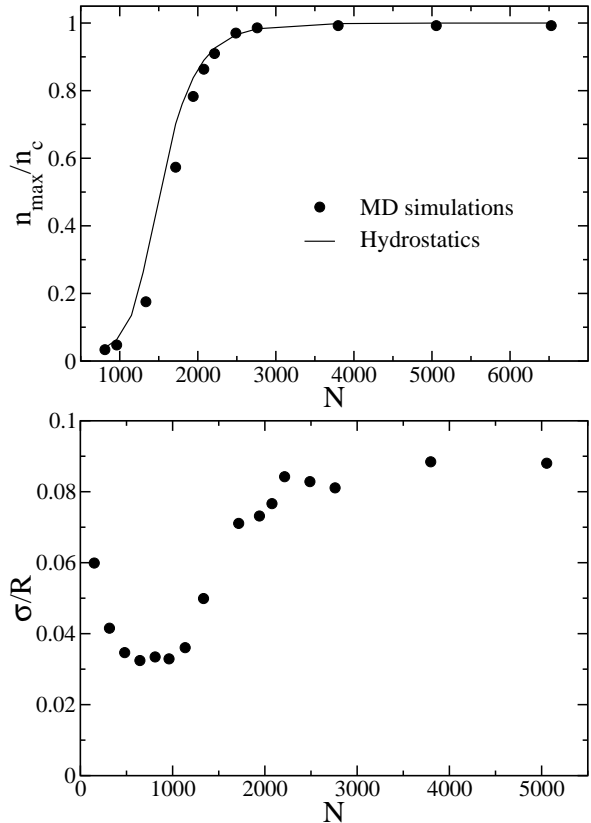

\includegraphics[width=0.9\columnwidth,clip]{fig3a.eps}
\includegraphics[width=0.9\columnwidth,clip]{fig3b.eps}
\caption{The top panel depicts the maximum scaled density
$n_{max}/n_c$ as a function of $N$ as predicted by the hydrostatic
theory with the CRs by Grossman \textit{et al.} \cite{Grossman} and
observed in MD simulations. The bottom panel shows the scaled
standard deviation of the center of mass position as a function of
$N$, obtained in MD simulations.} \label{fig3}
\end{figure}

Now let us compare the azimuthally symmetric density profiles, found from our
hydrostatic calculations, with the results of MD simulations. As noted above, in
the MD simulations the coefficient of normal restitution $\varepsilon=0.888$ was
fixed, while the number of particles $N$ was varied. Accordingly, the scaled
parameter $\Lambda =9980$  was fixed but $f$ was varied. Figure \ref{fig2} shows
a comparison of the hydrostatic radial density profiles with the radial density
profiles obtained in MD simulations, for four different values of $N$. The MD
profiles were averaged over 1000 snapshots. It can be seen from Fig. \ref{fig2}
that for $N=1716$ the theory overestimates the density in the center of the box.
This is expected as, for relatively small $N$, the maximum density is
considerably less than the close-packing density, and the accuracy of the CRs by
Grossman \textit{et al}. is not as good. For larger $N$ the agreement rapidly
improves. We also computed the density profiles using another set of
semi-empiric CRs: those obtained in the spirit of Enskog theory \cite{JR}. One
can clearly see from Fig. \ref{fig2} that the Enskog-type CRs predict
unphysically high densities in the cluster, and is also less accurate at
intermediate densities. Figure \ref{fig3} compares, at different $N$, the
\textit{maximum} radial densities predicted by hydrostatic theory with the CRs
by Grossman \textit{et al}. and those obtained in the MD simulations. The
agreement is very good for the densities approaching the close packing density.

\section{Fluctuations}

\begin{figure}
\centerline{\includegraphics[width=8.0cm,clip=]{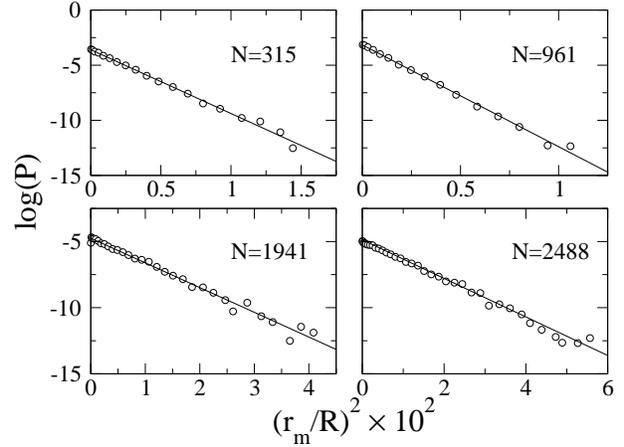}}
\caption{The logarithm of the radial PDF $P(r_m)$ of the center of
  mass position versus $(r_m/R)^2$, for four values of $N$.}
\label{fig4}
\end{figure}

Now we turn to stationary fluctuations of the radial coordinate
$r_m(t)$ of the center of mass of the system. The radial PDF $P(r_m,
t)$ is normalized by the condition
\begin{equation}
2\pi \int_0^R P(r_m, t)\, r_m dr_m =1\,,
\end{equation}
 where we have returned to the dimensional coordinate.
Typical MD results are presented in Fig. \ref{fig4}, which shows
$\log P(r_m)$ versus $(r_m/R)^2$, for four values of $N$.\footnote{For 
larger values of $N$ the density in the center of the container 
approaches the close packing (see Fig. \ref{fig2}), and the particle 
collision rate in the high-density region almost diverges. As a result, 
the event-driven simulation progresses extremely slowly, and the 
simulated real time is not large enough to allow for good statistics 
needed to determine a reliable PDF.} The
observed straight lines clearly indicate a Gaussian.
This finding suggests a Langevin description of the macro-particle.
Consider a macro-particle, performing an over-damped motion in a
confining harmonic potential $U(r_m)=k r_m^2/2$ and driven by a
(delta-correlated) discrete-particle noise $\eta(t)$. The Langevin
equation for this problem reads
\begin{equation}\label{Langevin}
h \dot{r}_m+k r_m=\eta(t)\,,
\end{equation}
where $h$ is the damping rate, $\langle \eta(t)
\eta(t^{\prime})\rangle=2h \Gamma \,\delta(t-t^{\prime})$, and
$\Gamma$ is an effective magnitude of the discrete particle noise.
In the limit $\Gamma \to 0$, the (deterministic) steady-state
solution of Eq. \eqref{Langevin} is $r_m=\dot{r}_m=0$: the
macro-particle at rest, located at the center of the box. At
$\Gamma>0$, the steady state PDF is (see, \textit{e.g.} Ref. \cite{Gardiner}) 
\begin{equation}
P(r_m)= \frac{1}{\pi \,\sigma^2}\,\exp \left(- \frac{r_m^2}{\sigma^2}\right)\,,
\end{equation}
 where
$\sigma^2 = \Gamma/k$, and the normalization constant is computed
for $\sigma \ll R$. The variance $\sigma^2$ is the ratio of
$\Gamma$ (a characteristic of the discrete noise) and $k$ (a
macroscopic quantity). Unfortunately, the present state of theory
does not enable us to calculate either of these two quantities. An
important insight, however, can be achieved from the
$N$-dependence of $\sigma$, obtained by MD. In analogy with
equilibrium systems, one might expect the relative magnitude of
fluctuations to decrease with increasing $N$. Surprisingly, this
is not what we observed, see the bottom panel of Fig. \ref{fig3}.
One can see that $\sigma(N)$ approaches a plateau, that is
fluctuations persist at large $N$.\footnote{The small-$N$ behavior
(the first three data points: $N=150$, $315$ and $480$)
in the bottom panel of Fig. \ref{fig3} agrees with the dependence
$\sigma/R = {\cal O}(N^{-1/2})$, expected for an ideal gas in
equilibrium. Not surprisingly, for these relatively small $N$ the
clustering effect is small. On the other hand, at very large $N$,
when the system approaches close packing (in our MD simulations,
this corresponds to $N = 36 275$), $\sigma/R$ must go to zero. We
could not probe this regime, however, as MD became prohibitevely
long already at $N \sim 15 000$.}

\section{Discussion}

We employed a simple model system \cite{Poeschel} to investigate the density
profiles and fluctuation properties of dense clusters emerging in driven
granular gases. The density profiles, obtained in our MD simulations, are well
described by hydrostatic equations which employ the constitutive relations
suggested by Grossman \textit{et al.} \cite{Grossman}. The good performance of
these constitutive relations is in agreement with previous results in rectangular
geometry, with and without gravity \cite{Grossman,MPB}. Marginal stability
analysis yields a natural explanation to the circular cluster shape, observed in
the MD simulations. Finally,  the observed Gaussian fluctuations of the center
of mass suggest an effective Langevin description in terms of a macro-particle
in a confining potential of hydrodynamic nature, driven by discrete particle
noise. The fluctuations persist as the number of particles in the system is
increased, and this surprising finding awaits a proper theoretical
interpretation.


\bibliographystyle{spbasic}
\bibliography{DMPS}   

\end{document}